\documentstyle[aps,floats,preprint,12pt]{revtex}

\input epsf

\def\BNL {Department of Physics, Brookhaven National Laboratory, Upton, NY 11973\\}
\def\WU {Department of Physics, Washington University, St. Louis, M0 63130 \\}

\newcommand{\BE}{\begin{equation}}
\newcommand{\EE}{\end{equation}}
\newcommand{\BEA}{\begin{eqnarray}}
\newcommand{\EEA}{\end{eqnarray}}
\newcommand{\EL}{\nonumber\\}

\newcommand{\etal}{{\em et al.\ }}

\newcommand{\gbeta}{6/g^2}

\newcommand{\op}{{\cal O}}
\newcommand{\kl}{\kappa_{l}}
\newcommand{\kh}{\kappa_{h}}
\newcommand{\Bd}{B^0}
\newcommand{\Bs}{B^0_s}
\newcommand{\MB}{M_{bd}}
\newcommand{\MS}{M_{bs}}
\newcommand{\MP}{M_{hl}}
\newcommand{\MC}{M_{hc}}

\newcommand{\CV}{1.76(10)^{+57}_{-42}}
\newcommand{\RDECAY}{1.17(2)^{+12}_{-6}}
\def\simge{
    \mathrel{\rlap{\raise 0.511ex
        \hbox{$>$}}{\lower 0.511ex \hbox{$\sim$}}}}
\def\simle{
    \mathrel{\rlap{\raise 0.511ex
        \hbox{$<$}}{\lower 0.511ex \hbox{$\sim$}}}}

\begin{document}

\title{{\small \rm \rightline{BNL-HET-98/7} \rightline{Wash.\ U.\ HEP/98-30} }
\vskip.5in
SU(3) Flavor Breaking in Hadronic Matrix Elements for $B - \bar B$ Oscillations }
\author{C. Bernard$^a$, T. Blum$^b$, and A. Soni$^b$}
\address{ $^a$\WU }
\address{ $^b$\BNL }
\date{\today}

\maketitle

\begin{abstract}
Results in the quenched approximation
for SU(3) breaking ratios of the heavy-light decay constants
and the $\Delta F=2$ mixing matrix elements 
are reported. Using lattice simulations at $6/g^2=5.7$,
5.85, 6.0, and 6.3, we directly compute the
mixing matrix element $M_{hl}=\langle \bar P_{hl}|\bar h \gamma_\mu
(1-\gamma_5)l \bar h \gamma_\mu(1-\gamma_5)l|P_{hl}\rangle$.
Extrapolating to the physical B meson states,
$B^0$ and $B_s^0$, we obtain $M_{bs}/M_{bd} = \CV$ 
in the continuum limit.  The systematic error includes the errors
within the quenched approximation but not the errors of quenching.
We also obtain the ratio of decay constants, $f_{bs}/f_{bd}=\RDECAY$.
For the B parameters we find $B_{bs}(2 GeV)=B_{bd}(2 GeV)=1.02(13)$;
we cannot resolve the SU(3) breaking effects in this case.
\end{abstract}
\newpage

Using lattice methods, one can calculate the $\Delta F=2$ heavy-light
mixing matrix element,
\BEA
\label{matrix hl}
\MP(\mu) \equiv \langle \bar P_{hl}|\bar h \gamma_\rho
(1-\gamma_5)l \bar h \gamma_\rho(1-\gamma_5)l|P_{hl}\rangle  \ .
\EEA
As is well known, these matrix elements govern $\Bd-\bar\Bd$ and $\Bs-\bar\Bs$
oscillations~\cite{FLYNN,SONI,BEES}. In the above $h$ and $l$ denote heavy and light quark
fields, $P_{hl}$ the corresponding pseudoscalar meson, and $\mu$ is the energy 
scale appropriate to the calculation. 
Here we compute directly the SU(3) flavor breaking ratio~\cite{US96},
\BEA
\label{ratio}
r_{sd}=\MS(\mu)/\MB(\mu)\ .
\EEA 

Our central result is that $r_{sd}=\CV$ in the 
quenched approximation, where the first error is
statistical and the second systematic.
The importance of this ratio is that, in conjunction with the 
eventual experimental measurement
of $\Bs-\bar\Bs$ oscillations, it should allow the 
cleanest extraction of the crucial CKM parameter $V_{td}$. 

Since 
the
CKM matrix elements are fundamental parameters of the Standard Model, 
it is clearly important to determine them precisely. $V_{td}$ is 
especially significant because
low energy manifestations of
$CP$ violation, which enter through virtual $t$--$\bar t$ loops,
invariably involve $V_{td}$. 
At present, $V_{td}$ is deduced
from $\Bd-\bar\Bd$ oscillations via
the mixing parameter $x_{bd}=\Delta M_{bd}/\Gamma_{bd}$\cite{BURAS}.
\BEA
\label{mixing par}
x_{bd} &=&\tau_{bd} \frac{G_F^2}{6 \pi^2}m_{bd}
b(\mu) B_{bd}(\mu)f_{bd}^2 M_W^2 \eta_{QCD} S(x_t) |V_{td}|^2,
\EEA
where $m_{bd}$, $\tau_{bd}\equiv\Gamma^{-1}_{bd}$, and $f_{bd}$ 
are the mass, life time, and
decay constant of the $\Bd$ meson, and $(\Delta M)_{bd}$ is the mass difference of the
two mass eigenstates of the $\Bd-\bar\Bd$ system. $x_{bd}$ is the mixing parameter 
characterizing the oscillations and has been determined experimentally, 
$x_{bd}=0.73(5)$ ~\cite{PDB}. 
$B_{bd}$ is the so called bag parameter,
and $b(\mu)$ and $S(x_t)$ are perturbatively calculated short 
distance quantities\cite{BURAS}.
To extract $V_{td}$ from Eq.~(\ref{mixing par}) 
requires knowledge of two hadronic
matrix elements, $f_{bd}$ and $B_{bd}$. 
These are being calculated using lattice and other methods.
$f_{bd}$ may eventually be measured experimentally through,
for example, the decay $B\to \tau \nu_\tau$. 
However, $B_{bd}$ is a purely theoretical
construct which is inaccessible to experiment. 
Thus determination of $V_{td}$ from
experiment will ultimately be limited by the precision of the 
nonperturbative quantity
$f^2_{bd}B_{bd}$. These parameters are related to the matrix element
Eq.~(\ref{matrix hl}) via
\BEA
\label{B param}
\MB(\mu)=\frac{8}{3}f_{bd}^2 m_{bd}^2 B_{bd},
\EEA
and often one writes $b(\mu)M_{bd}(\mu) = \hat M_{bd}$, a
renormalization group invariant (RGI) quantity.

Making the replacement $d\to s$ in Eq.~(\ref{mixing par}) and taking the
ratio with Eq.~(\ref{mixing par}), we arrive at an alternate way to 
extract $V_{td}$, 
\BEA
\label{ckm ratio}
\frac{|V_{td}|^2}{|V_{ts}|^2}= r_{sd} 
\frac{m_{bd}}{m_{bs}}
\frac{\tau_{bs}}{\tau_{bd}}
\frac{x_{bd}}{x_{bs}}
\EEA
Thus, in contrast to the above method for determining $V_{td}$ via use of 
Eq.~(\ref{mixing par}), once the $\Bs-\bar\Bs$ oscillation parameter, 
$x_{bs}\equiv\frac{(\Delta M)_{bs}}{\Gamma_{bs}}$, is experimentally measured,
we can use Eq.~(\ref{ckm ratio}) to determine $V_{td}$.
The right hand side of Eq.~(\ref{ckm ratio}) involves three SU(3) 
breaking ratios, only one
of which, namely $r_{sd}$, needs to be calculated non-perturbatively. 
The remaining two can be measured experimentally, 
at least in principle.  
Indeed, since the spectator approximation is expected to hold to
a very high degree of accuracy \cite{BIGI}, it is also reasonable
to expect that $\tau_{bs}/\tau_{bd}=1$ within a few percent.
Of course, the measurement of $x_{bs}$ is very challenging. 
A variety  
of experimental efforts are underway at both $e^+e^-$ and hadronic
machines towards that goal \cite{LEPBs}. 
Note also that $V_{ts}$ in Eq.~(\ref{ckm ratio}) is related
by three generation unitarity to $V_{cb}$ and is therefore already quite
well determined, $|V_{ts}|\approx|V_{cb}|=0.041 \pm 0.003\pm .002$\cite{PDB}.
The important distinction between using Eq.~(\ref{ckm ratio}) instead of 
Eq.~(\ref{mixing par}) is that the former requires only 
knowledge of {\it corrections}
to SU(3) flavor symmetry while the latter 
requires the {\it absolute} value of the
matrix element $\MB$.
It is also important to realize that since $r_{sd}$ is a ratio of two 
very similar hadronic
matrix elements, it is 
less susceptible to common systematic 
errors in lattice 
calculations, among which are scale dependence, 
matching of continuum and lattice
operators, and heavy quark mass dependence. Indeed, the ratio
$r_{sd}$ is, to an excellent approximation, RG 
invariant, even though the individual matrix elements
$M_{bs}$ and $M_{bd}$ are scale dependent.

In passing, we 
recall that 
flavor symmetries have also played a 
crucial role in determining other CKM matrix elements.
In particular,
SU(3) flavor symmetry has been important in precisely determining 
$V_{us}\equiv \sin \theta_c$. 
More recently, heavy quark symmetry (HQS)\cite{MBW} has been used to
improve systematically the determination of $V_{cb}$.

The lattice methodology for calculating these matrix elements ({\it i.e.} 
Eq.~(\ref{matrix hl}))
is, by now, well known~\cite{CB}. The amplitudes
for $B_0-\bar B_0$ mixing, usually called ``box" amplitudes, occur at second order in the 
weak interaction. After integrating out the W boson, the operator product expansion (OPE)
allows one to write the corresponding amplitude as a short distance expansion.
In this case there is only one operator in the expansion,
${\cal O}_{LL}=\bar b \gamma_\mu(1-\gamma_5)d \bar b \gamma_\mu(1-\gamma_5)d$.
Its (Wilson) coefficient, $C_{LL}(\mu)$, is calculated most easily in continuum
perturbation theory.
The matrix element of $\op_{LL}$ 
must be calculated non-perturbatively on the lattice since it
contains the long distance QCD information of the physical process in 
question. The product of the two yields the scale invariant amplitude, 
which is obtained by translating either result from 
one regularization scheme to the other. We accomplish this 
in the usual way by matching the lattice 
operator to the continuum operator in a particular scheme at some low energy scale.
For convenience we choose the scale $\mu=2$ GeV.
Using the renormalization group equations, $C_{LL}(M_W)$ 
is then run down to this scale, which yields $C_{LL}(\mu)$.

For Wilson quarks the continuum-lattice matching for ${\cal O}_{LL}$ has been 
carried out to one loop in perturbation theory~\cite{MART,BDS,GUPetal}.
\BEA
\op_{LL}^{cont} = 4{\tilde\kh\tilde\kl}\left(\op_{LL}^{latt} + 
\frac{g^2}{16 \pi^2}\left(\right. \right. Z_{+}(a\mu) \op_{LL}^{latt} \EL
+ \left.\frac{Z^{\ast}}{48}(2\op_{SS}
+ 6\op_{PP}-11\op_{VV} + 11\op_{AA} +2\op_{TT}))\right)
\label{match eq}
\EEA
where $\op_{ii}$ 
corresponds to $\gamma_\mu(1-\gamma_5)\to$ 1, $\gamma_5$, $\gamma_\mu$, 
$\gamma_5 \gamma_\mu$, and $\sigma_{\mu\nu}$ in the expression for $\op_{LL}$. 
The Wilson quark action explicitly breaks chiral symmetry, so 
these new operators arise to cancel the chiral symmetry breaking terms in 
$\op_{LL}^{latt}$. We use the naive dimensional regularization (NDR) scheme
with ``tadpole improvement,'' so
$Z_{+}=(-50.841-4\ln{(a\mu)+34.28)}$~\cite{GUPetal,MART,BDS} where $a$ is the 
lattice spacing.
$Z^{\ast}=9.6431$ and depends only on the Wilson r parameter which we set to 1.
The last term in $Z_{+}$ comes from mean field improved perturbation 
theory~\cite{LM},
which removes tadpole terms. ($Z^*$ is an off-diagonal
correction which does not have tadpole contributions at this order.)
The scale at which the coupling $g$ in Eq.~\ref{match eq}
is to be evaluated is not fixed at one loop, however.
It has been estimated for the decay constant
using the methods of Ref.~\cite{LM} as
$q^{*}=2.316/a$ \cite{BGM}.
We use this scale to find the central values;
the variation with two choices for the scale, $1/a$ and $\pi/a$, 
is used to determine the associated systematic error.
The usual naive renormalization of the fermion fields,
$4{\kh\kl}$, is 
modified by the El-Khadra-Kronfeld--Mackenzie(EKM) norm~\cite{EKM} 
which is more suitable for the heavy quarks in our simulations.

The Wilson quark action also introduces errors proportional to (powers of) 
the lattice spacing in observables. 
We attempt to remove these by simulating at several values of the coupling 
$6/g^2(a)$ and extrapolating to $a=0$.

Table~(\ref{lat table}) summarizes 
the lattice data used in our analysis. For each $\kl$ and 
$\kh$ in Table~\ref{lat table} we calculate a quark propagator 
using a single point source at the center of the lattice and
a point sink. These are contracted to obtain two and three point 
meson correlation functions which are fit simultaneously to obtain
the matrix element $\MP$.

\begin{table*}
\caption{Summary of simulation parameters. $\kappa_h$ and $\kappa_l$
are the heavy and light Wilson quark hopping parameters.}
\label{lat table}
\begin{tabular}{ccccc}
$\gbeta$ & conf. & size & $\kappa_{light}$ &$\kappa_{heavy}$ \\
\hline
5.7&	83&$	16^3\times 33$&	0.160 0.164 0.166&0.095 0.105 0.115 0.125 0.135 0.145\\
5.85&	100&$	20^3\times 61$&	0.157 0.158 0.159 0.160&0.092 0.107 0.122 0.130 0.138 0.143\\
6.0&	60&$	16^3\times 39$&	0.152 0.154 0.155&0.103 0.118 0.130 0.135 0.142\\
6.0&	100&$	24^3\times 39$&	0.152 0.154 0.155&0.103 0.118 0.130 0.135 0.142\\
6.3&	100&$	24^3\times 61$&	0.148 0.149 0.150 0.1507&0.100 0.110 0.125 0.133 0.140\\
6.5&	40&	$32^3\times 75$&0.146 0.147 0.148 0.1486&0.100 0.110 0.120 0.132 0.137 0.142\\
\end{tabular}
\end{table*}

\begin{figure}[hbt]
\vspace{-.25in}
    \vbox{
        \epsfxsize=6.0in \epsfbox[0 0 4096 4096]{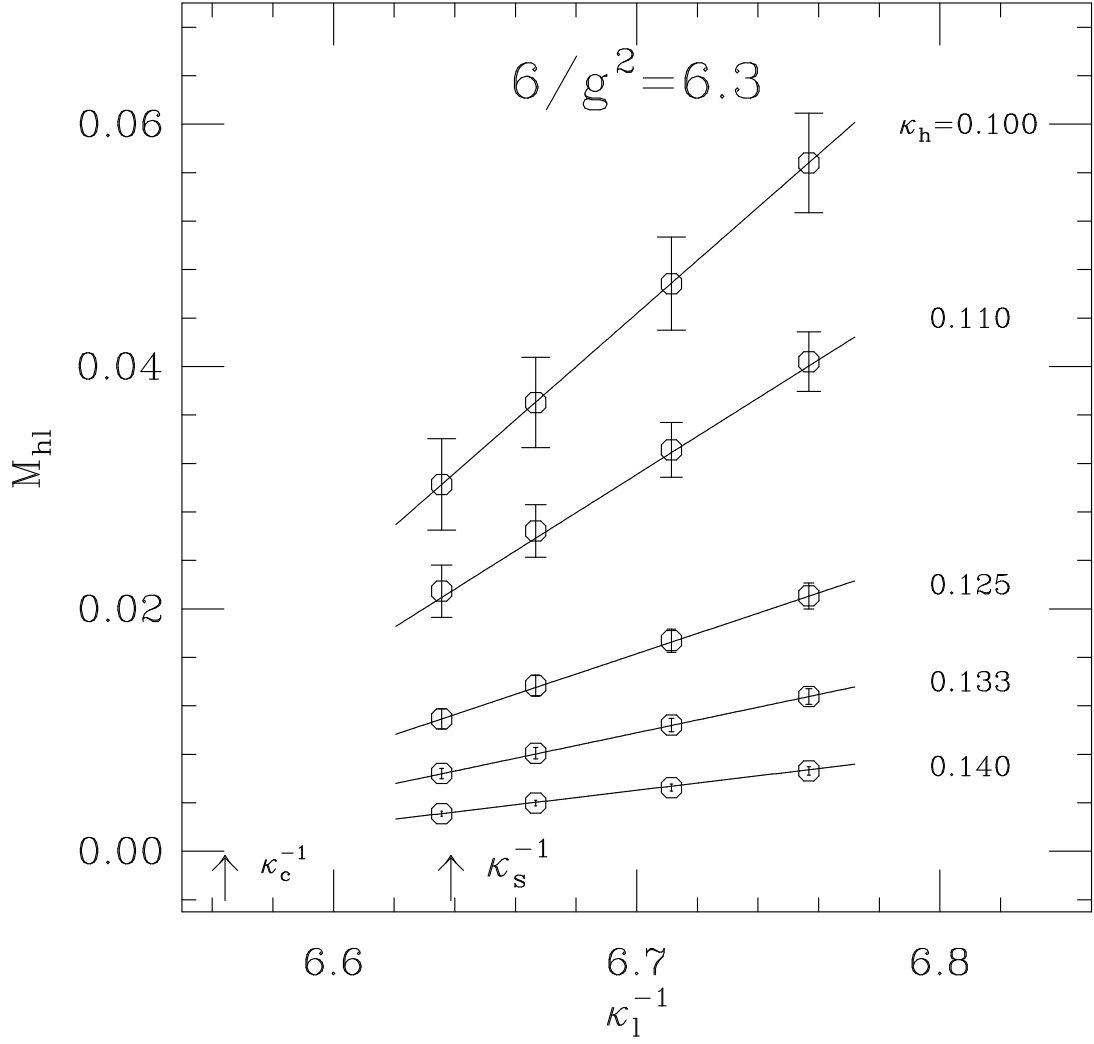}
    }
    \caption{\label{mllc} The four quark matrix element $\MP$ at $\gbeta=6.3$
	as a function of the inverse light quark hopping parameter.
        Results for the other values of $\gbeta$ are similar. 
	The solid lines are covariant linear fits to the data.}
\end{figure}

In Fig.~\ref{mllc} we show sample results at $\gbeta=6.3$ for $\MP$ 
{\em vs.} $\kl^{-1}$ for each value of $\kh$,
where the quark mass in units of the lattice spacing 
is $am_q=(\kappa_q^{-1}-\kappa_c^{-1})/2$. 
Here, $\kappa_c$
is the critical hopping parameter where the pion mass vanishes.
$\MP$ is extracted
from the three point pseudoscalar correlation function, which
is proportional to $\MP$ for large time separations of the
four quark operator, and the two pseudoscalar meson interpolating operators. 
Results for the physical $B$ and $B_s$ meson systems follow from
a series of fits to the lattice data, which we use to extrapolate in the
two parameters $\kh$ and $\kl$. We use covariant fits and a jackknife 
procedure at each step to account for 
the correlations in the data. 

\begin{figure}[hbt]
\vspace{-.25in}
    \vbox{
        \epsfxsize=6.0in\epsfbox {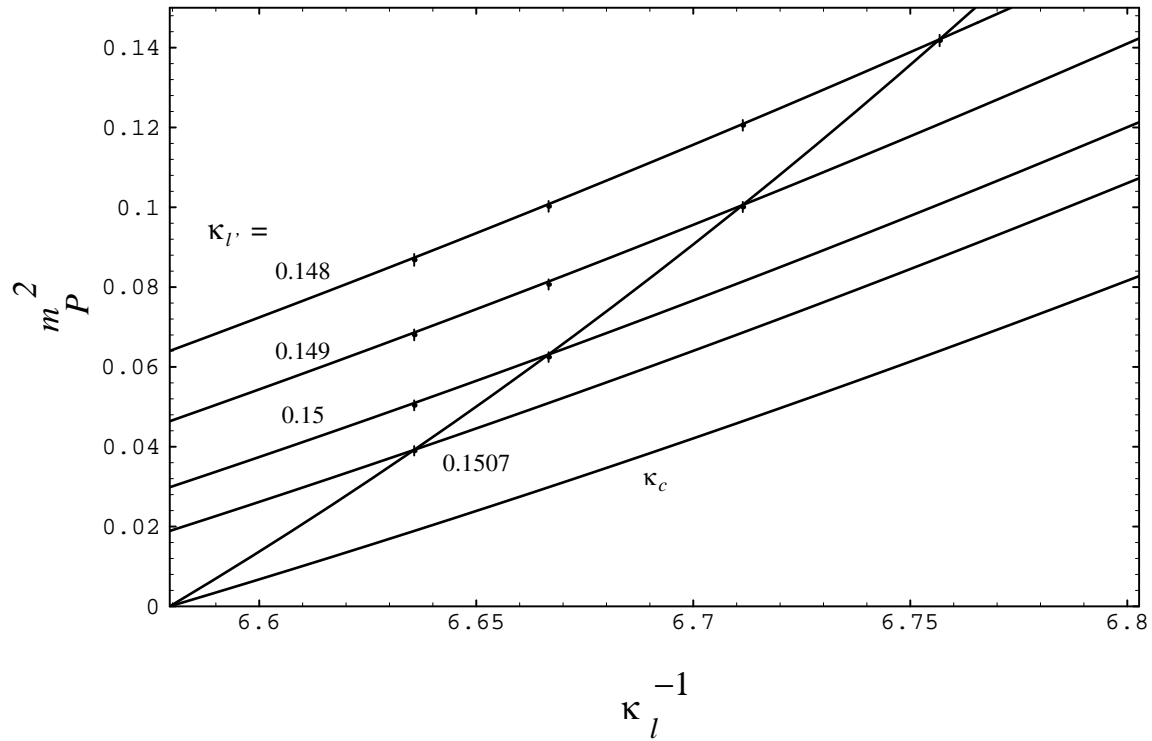}
    }
    \caption{\label{mpi2} The pseudoscalar mass squared as a 
	function of the non-degenerate light quark hopping parameters. 
	$\gbeta=6.3$.
        Results for the other values of $\gbeta$ are similar. 
	The solid lines are from a covariant fit 
	to the form in Eq.~\protect\ref{kappa_c fit}.} 
\end{figure}

To begin the extrapolations, $\kappa_c$ and $\kappa_s$ (the strange
quark hopping parameter) are determined
from a fit to the squares of the pseudoscalar 
masses as a function
of $\kl^{-1}$ and $\kappa_{l^\prime}^{-1}$ ($l$ and $l^\prime$ refer to 
non-degenerate light quarks ). We use the following fit form,
which does {\it not} include the logarithmic terms relevant at very
small quark mass~\cite{BG}.
\BEA
\label{kappa_c fit}
m_{ll^\prime}^2&=&c_0 + c_1 (\kappa_l^{-1} + \kappa_{l^\prime}^{-1}) 
+ c_2 \kl^{-1} \kappa_{l^\prime}^{-1}
+ c_3 (\kappa_l^{-2} + \kappa_{l^\prime}^{-2}).
\EEA
A typical fit is shown in Fig.~\ref{mpi2}.
The values for $\kappa_c$ and $\kappa_s$ and $\chi^2/$dof for each fit
are summarized in Table~\ref{kctable}. The curvature in Fig.~\ref{mpi2}
is small but certainly present:
including only constant and linear
terms in the fits generally yields poor $\chi^2$ values. 
The linear fits shown in Table~\ref{kctable} were obtained by omitting the 
3, 2, 0, 3, 8, and 6 heaviest
points for $\gbeta=5.7$ to 6.5, respectively.
The linear fits then had acceptable values of $\chi^2$ except
at $\gbeta=6.0(24^3)$ and 6.3 where completely constrained fits were 
used. 
The values for $\kappa_c$ 
obtained from the linear fits are in
rough agreement with the quadratic fits; they are systematically low
by one to two statistical standard deviations.
Values for $\kappa_c$ determined from the linear fits
agree with earlier calculations~\cite{QCDPAX,BGKS,APE} at $\gbeta=6.0$ and
\cite{QCDPAX} at $\gbeta=5.85$.
At $\gbeta=5.7$, 5.85, 6.0, and 6.3, $\kappa_c$ is systematically
higher by several statistical standard deviations than the values
found by the MILC collaboration~\cite{CLAUDE}, and the value at
6.0 in Ref.~\cite{GOCK}. In this study we use point sources on lattices with
modest extent in the time direction. A detailed comparison with the data 
from Ref.~\cite{CLAUDE} indicates that this is likely to be
the main cause of the discrepancy.

Since higher order chiral effects are completely different in the
quenched and full theories, one might argue that the linear fits
are preferable on physical grounds. For our central values, we stick with
the quadratic fits, which describe our data well, but we take the difference
arising from a switch to linear fits (as well as from the $\kappa_c$
shift necessary to reproduce the Ref.~\cite{CLAUDE} data) as 
an estimate of one source of systematic errors.

Finding $\kappa_s$ requires
the scale $a$, which we set from $af_\pi$, to determine the lattice 
value of the kaon mass $a m_K$ ($a^{-1}$ is also tabulated in 
Table~(\ref{kctable})). 
Our values for $\kappa_s$ using linear fits 
agree to about one $\sigma$ with Ref.~\cite{QCDPAX} at $\gbeta=5.85$ and
Refs.~\cite{QCDPAX,BGKS} at 6.0. Refs.~\cite{QCDPAX,BGKS} 
used $m_\rho$ to set the lattice spacing and $am_K$, among others,
to determine $\kappa_s$. Here we compare with values determined from
$am_K$. At $\gbeta=5.7$, 5.85, 6.0,
and 6.3, $\kappa_s$ determined from $am_K$ (with $af_\pi$ used
to set the lattice spacing) agrees well with the results
from Ref.~\cite{CLAUDE}. 
One might expect the values of $\kappa_s$, like $\kappa_c$,
 to disagree 
among the various calculations since 
they are determined from the same data.  However, the added 
statistical uncertainty
from the kaon mass is enough to mask the systematic error.
We mention the above
because the flavor breaking ratios given below are sensitive to the 
(relative) values of $\kappa_c$ and $\kappa_s$.
We also note that at $\gbeta=5.7$ the choice of the coupling constant scale
for $Z_A$, the lattice axial current renormalization which appears in the 
determination of $f_\pi$, has a significant effect on the 
lattice spacing determination; 
$Z_A$ differs by $\sim 7\%$ when the scale changes from
$1/a$ to $\pi/a$.

\vskip .7truein
\begin{table}[hbt]
\caption{Inverse lattice spacing and critical and strange hopping parameters.
For each value of $\gbeta$, the two rows correspond to  a determination
of $\kappa_c$ and $\kappa_s$ by quadratic and 
linear fits, respectively, to the pseudoscalar spectrum.
For the linear fits, the 3, 2, 0, 3, 8, and 6 heaviest
points are omitted for $\gbeta=5.7$ to 6.5, respectively. 
Each value of $\kappa_c$ results in a
corresponding lattice spacing from $a f_\pi$.
$\chi^2$/dof refers to the fit used to determine the quantity immediately
to the left. 
An entry of ``cf" means a completely constrained fit.}
\label{kctable}
\begin{tabular}{ccccccc}
$\gbeta$&  $a^{-1}$(GeV)&$\chi^2/{\rm dof}$&$\kappa_c$& $\chi^2/{\rm dof}$& $\kappa_s$ & $\chi^2/{\rm dof}$\\
\hline
5.7&    1.37(10)&0.31/2&       0.16973(15)&	0.24/2&	0.1645(7)&	0.24/2\\
   &    1.35(9)&0.31/2&       0.16953(9)&	0.43/1&	0.1640(8)&	0.43/1\\
5.85&   1.65(13)&0.01/1&       0.16170(8)&	0.33/3&	0.1576(6)&	0.33/3\\
    &   1.64(13)&0.01/1&       0.16157(5)&	1.30/4&	0.1575(9)&	1.30/4\\
6.0($16^3$)&    2.03(17)&3.16/1&       0.15725(23)&	0.62/1&	0.1544(5)&	0.62/1\\
           &    2.01(16)&3.16/1&       0.15715(6)&	1.08/3&	0.1545(4)&	1.08/3\\
6.0($24^3$)&    2.08(13)&0.67/1&       0.15714(4)&	2.6/1&	0.1544(4)&	4.0/1\\
           &    2.17(15)&0.67/1&       0.15739(4)&	cf&	0.1548(4)&	cf\\
6.3&    3.09(21)&0.81/2&       0.15199(4)&	9.5/6&	0.1502(2)&	9.5/6\\
   &    3.10(21)&0.81/2&       0.15191(4)&	cf&	0.1503(2)&	  cf\\
6.5&    4.29(50)&0.38/2&       0.14993(18)&	1.04/5&	0.1486(3)&	1.04/5\\
   &    4.22(49)&0.38/2&       0.14972(1)&	3.04/1&	0.1487(3)&	3.04/1\\
\end{tabular}
\end{table}

Next, we linearly extrapolate $\MP$ to $\kl=\kappa_c$ and $\kappa_s$. 
The results for $\MP$
at $\gbeta=6.3$ (see Fig.~\ref{mllc}) show a smooth linear behavior. 
Similar results are obtained at the other couplings. 
Up to this step all of the covariant 
fits have acceptable values of $\chi^2$, except 
the point at $\gbeta=6.0$ ($24^3$). Results at this point also showed
significant variation with the form of the chiral extrapolation. The three point
correlators here do not exhibit true plateaus but instead monotonically
decrease with time, so there is undoubtedly
contamination from excited states and additional
uncertainty coming from the choice of fit range, which is necessarily small. 
Also, at $\gbeta=6.5$ the data were too noisy to extract $\MP$. 
Finally, the point at $\gbeta=5.85$ is somewhat problematic. 
The statistical errors are large,
so this point does not have a large impact on the continuum extrapolation.	
The difficulty arises in the three point correlators which show plateaus
with a somewhat large oscillation. Three of the four light $\kappa$'s happen to
be below $\kappa_s$ while the fourth is just above. Thus, our light
$\kappa$'s are closely spaced. The above considerations
lead to a relatively inaccurate determination of 
the slope of $\MP$ {\em vs.} $\kl^{-1}$,
which essentially determines $r_{sd}$.

The heaviest mass points in our calculation suffer from heavy quark 
systematic errors;
the lattices are too coarse to resolve objects with mass greater than 
the inverse lattice spacing. 
The biggest correction of these errors comes from using the EKM norm 
mentioned above. An additional
correction can be made by using the so-called kinetic 
mass~\cite{EKM} in place of the meson (pole) mass in the
heavy mass extrapolations described below. 
As in Ref.~\cite{BLS}, we use the tadpole improved tree level definition of 
the kinetic mass, $m_P^{kin}= m_P + 
\tilde m_2 - \tilde m $.
$\tilde m_2$ and $\tilde m$ are the tadpole improved heavy quark kinetic
and pole masses, respectively. This definition is motivated by a non-relativistic
expansion of the heavy-light meson mass and reduces to the usual 
meson pole mass in the limit where the heavy quark becomes light. 
For the heaviest masses, the kinetic pseudoscalar mass is almost double
the pole mass. This correction is also used in Refs.\cite{MILC,JLQCD}.

\begin{figure}[hbt]
\vspace{-.25in}
    \vbox{
        \epsfxsize=6.0in \epsfbox[0 0 4096 4096]{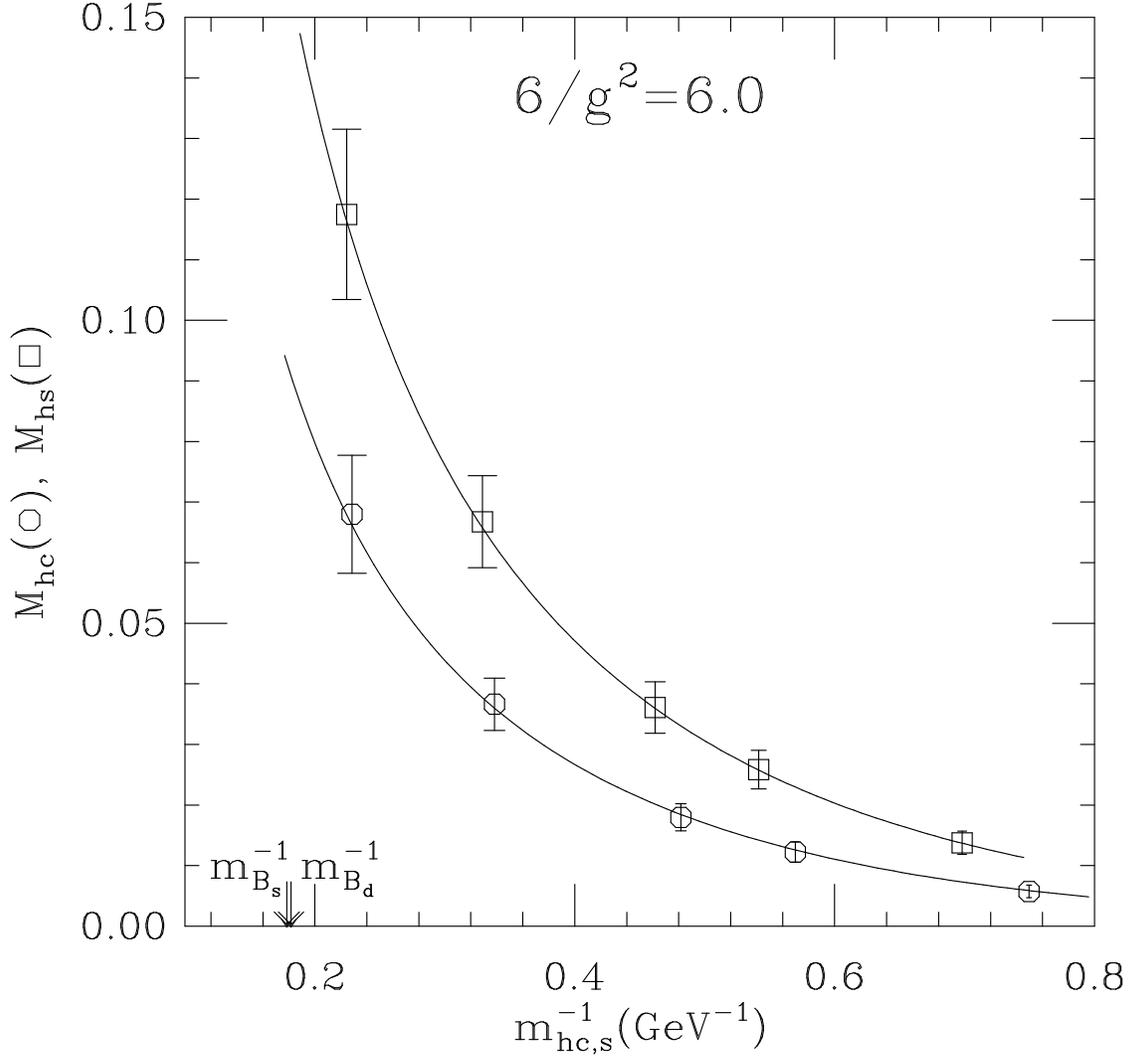}
    }
    \caption{\label{Mvsminv} The four quark mixing matrix element for down(octagons)
	and strange(squares) pseudoscalar mesons as a
        function of the inverse heavy-down(strange) meson mass at 
	$\gbeta=6.0$($16^3$). The solid lines are covariant fits using
	Eq.~\protect\ref{heavy fit} to 
	all of the data points. We find very similar results at the other  
	values of $\gbeta$.
    }
\end{figure}

We fit $\MC$ to the HQET form 
\BEA
\label{heavy fit}
\MC &=& c_{-1}m_P + c_0 + c_1 \frac{1}{m_P}.
\EEA
Here $m_P$ is any definition of the heavy-light pseudoscalar
mass. The resulting
fit is evaluated at the experimentally known $B^0$ meson mass 
to determine the physical value of the matrix element.
For the heavy-strange case we first extrapolate $\MP$ to
$\kappa_s$ instead of $\kappa_c$. The form in Eq.~\ref{heavy fit} follows
from the HQET results for the decay constants~\cite{FLYNN,SONI,MBW}
and the $B$ parameters:
\BEA
f_{P}\sqrt{m_P}&=& d_0 + d_1/m_P + O(m_P^{-2}),
\label{heavy decay} 
\EEA
\BEA
B_{P}&=&  b_0 + b_1/m_P + O(m_P^{-2}).
\EEA
Our data are consistent with these forms.
We note that for each value of $\gbeta$ 
all of the data points are covariantly fit to the above form,
and each fit has a good confidence level except the one at
6.0 ($24^3$). An example is shown in Fig.~\ref{Mvsminv}. 
It is noteworthy that the data fit the form in Eq.~\ref{heavy fit}
over such a large range (this is true for all of the couplings we studied).
At each coupling the heaviest (kinetic) mass is close to
the physical $B$ mass. 

\begin{table*}
\caption{Summary of results for $r_{sd}$.
The last two rows refer to constant and linear continuum
extrapolations, respectively. 
The errors shown in parentheses are statistical. 
Column 1 gives central values;
columns 2-8 represent systematic differences in $r_{sd}$
relative to column 1 and are used to estimate the 
corresponding error (see text).}
\label{r_sd table}
\begin{tabular}{ccccccccc}
$\gbeta$&1&2&3&4&5&6&7&8 \\
\hline
5.7&	        1.65(15)&1.64 (15)&1.50 (15)&1.52 (25)  &1.52 (25)&1.68 (15)&1.59 (14)&1.76 (20)\\
5.85&	   	1.78(25)&1.75 (24)&1.44 (30)&2.87 (2.20)&2.07 (45)&1.77 (27)&1.68 (21)&2.21 (27)\\
6.0$(16^3)$&	1.80(23)&1.79 (23)&1.64 (25)&1.70 (23)  &1.76 (22)&1.74 (19)&1.66 (19)&1.79 (23)\\
6.3&		1.96(23)&1.94 (23)&1.95 (33)&1.85 (34)  &1.97 (26)&1.83 (19)&1.68 (16)&1.99 (30)\\
$\infty$&	1.76(10)&1.75 (10)&1.58 (8) &1.67 (15)  &1.77 (13)&1.74 (9) &1.64 (8) &1.90 (12)\\
$\infty$&	2.18(39)&2.21 (42)&2.03 (42)&2.10 (56)  &2.29 (48)&1.94 (34)&1.76 (29)&2.09 (49)\\
\end{tabular}
\end{table*}

\begin{figure}[hbt]
\vspace{-.5in}
    \vbox{
        \epsfxsize=6.0in \epsfbox[0 0 4096 4096]{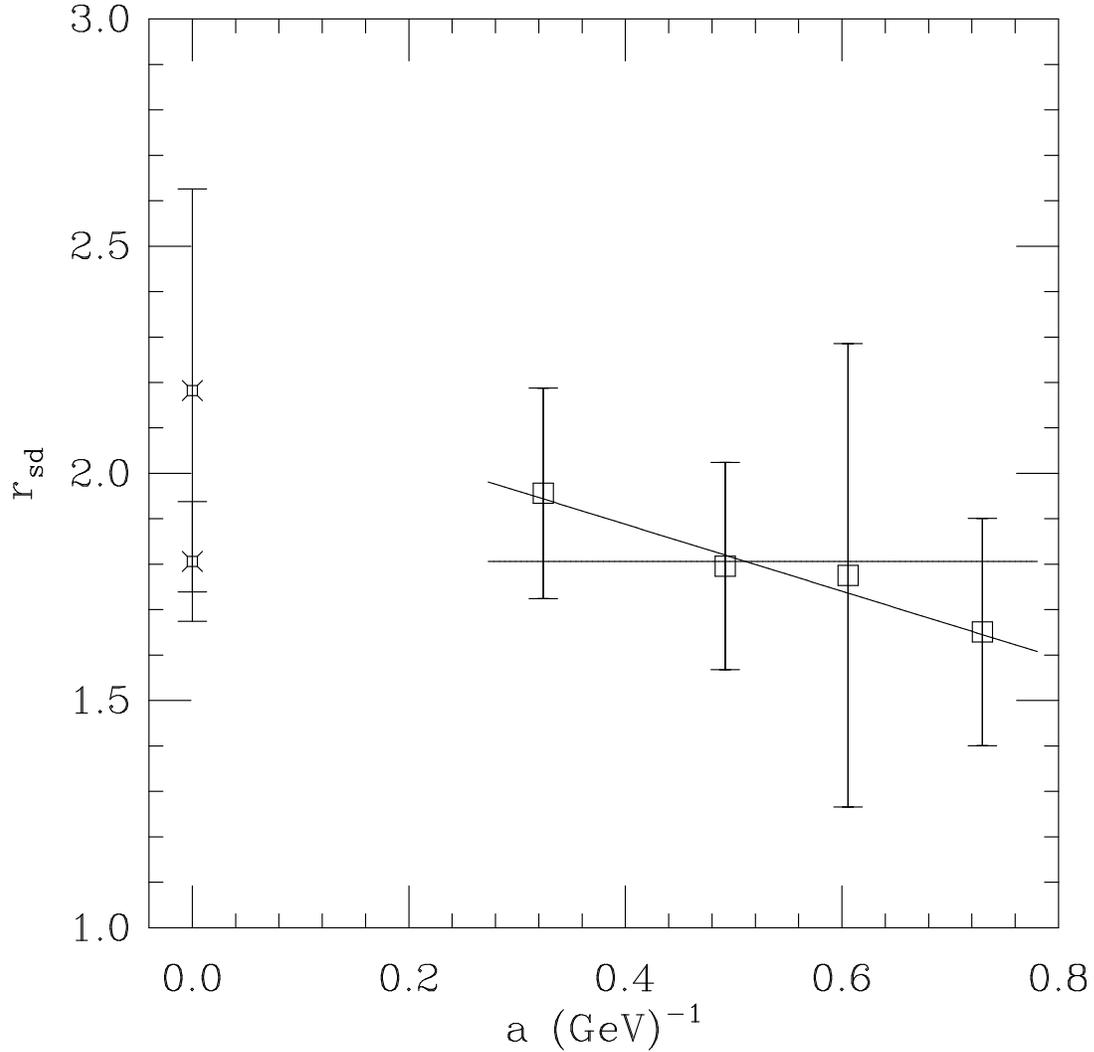}
    }
       \caption{\label{mat ratio} The SU(3) flavor breaking ratio $r_{sd}\equiv \MS/\MB$
               {\em vs.} the lattice spacing $a$. $\mu=2$ GeV and
		the coupling in Eq.~\protect\ref{match eq} has been evaluated
		at $q^{*}$.
		The lines denote constant and linear fits to the data, fancy 
		squares the corresponding continuum extrapolations.}
\end{figure}

Fig.~(\ref{mat ratio}) shows $r_{sd}=\MS/\MB$ as a function of $a$. 
The ratio is greater than unity for each value of $\gbeta$.
Using Eq.~\ref{match eq}, the renormalization scale is set to $\mu=2$ GeV and 
the coupling is evaluated at $q^{*}$. 
$r_{sd}$ is also tabulated in Table~\ref{r_sd table}.

As mentioned earlier, we expect the Wilson quark action to introduce
discretization errors of order $a$ in all observables.
However, for the ratio of two similar
quantities, we also expect a
significant cancellation of these
errors. A constant fit gives $\MS/\MB= 1.76(10)$
while a linear extrapolation in $a$ gives 2.18(39) (column 1 in Table
~\ref{r_sd table}). The above fits
have small $\chi^2$ values due to the large statistical errors, and
we cannot rule out one fit in favor of another based on $\chi^2$.
The measured slope for the linear fit differs from zero by $\simle 1\,\sigma$. 
The linear trend may easily
disappear with a one standard deviation change in either of the two
end points, so we use the constant fit as our central value and
the linear result as an estimate of the systematic error in the
continuum extrapolation.

Next we estimate other systematic uncertainties in our analysis.
The details are given in Table~\ref{r_sd table}. 
Columns 2-8 refer to separate analyses where one parameter was changed from its
reference value used to obtain column 1. The difference in the new 
extrapolated value is then taken as an estimate of the systematic error in $r_{sd}$.
In the following we list the uncertainties (numbers in parentheses refer to the corresponding
column in Table~\ref{r_sd table}).
(2.) Changing the coupling constant scale to $a^{-1}$ yields $r_{sd}=1.75(10)$ and
2.21(42) for constant and linear continuum extrapolations, respectively. 
(3.) Using quadratic chiral extrapolations for the matrix elements
yields 1.58(8) and 2.03(42). 
The fits used for the chiral extrapolations were completely constrained 
except at $\gbeta=6.3$ and 5.85, each of which had one degree of freedom.
(4.) Using the uncorrected pole mass yields $r_{sd}= 1.67(15)$ and
2.10(56). For this last case
we note that $\chi^2$ values were uniformly poor. 
The resulting fits underestimated the data at the heavy masses; the
lighter masses, which had smaller statistical errors, dominated the fits.
(5.) Constrained fits using only the heaviest masses 
yield 1.77(13) and 2.29(48), which we use to estimate the systematic error from
including heavy masses that may be too light.
(6.) As previously noted, we expect $r_{sd}$ to be sensitive to small relative
shifts between $\kappa_c$ and $\kappa_s$. 
Using the linear extrapolations for $\kappa_c$ and $\kappa_s$, we find
1.74(9) and 1.94(34).
The main effect is to lower the value of $r_{sd}$ at $\gbeta=6.3$,
which primarily affects the linear continuum extrapolation. 
(7.) As mentioned earlier, our values of $\kappa_c$ may be systematically high.
Shifting $\kappa_c$ by -0.0003 at each
coupling gives $r_{sd}=1.64(8)$ and 1.76(29). 
The shift was estimated
from the difference of our $\kappa_c$ values with
those of Ref.~\cite{CLAUDE} where Gaussian sources and longer lattices
in the time direction were used.
While the absolute shift is numerically small, it amounts to several
statistical standard deviations and is thus not accounted for in the
jackknife analysis.

(8.) Finally, we estimate the systematic error resulting from changing
the fit range of the three point correlation functions. 
Thus far acceptable values of $\chi^2$ were obtained using covariant fits
to the three point correlation functions. However, with point sources
(and lattices with modest extent in the time direction) the correlators do
not exhibit long plateaus, so the allowed fit range is necessarily
small. The fit ranges were shifted up or down by one or 
two time slices at each $\gbeta$ which generally resulted in worse
$\chi^2$ fits.
The only appreciable variations were at $\gbeta=5.7$
and 5.85. 

Columns 2,3,4,6, and 7 give lower results than the central value (column 1)
for the preferred (constant) continuum extrapolation.  Adding the differences
linearly gives a systematic error estimate of $-0.42$.  Combining the positive
differences in columns 5 and 8 with that from the linear continuum
extrapolation gives $+0.57$. The final result is then $r_{sd}=\CV$.


After our initial determination\cite{US96},
another group calculated $r_{sd}$ in the 
{\it static} approximation at $\gbeta=6.0$.
They find 1.35(5)\cite{GIM}.
When extrapolated to the static limit, $m_P^{-1}=0$, our data at $\gbeta=6.0$ 
yield $r_{sd}=1.39(30)$ which agrees well with the above. In addition,
our data at each value of $\gbeta$ indicates that
$r_{sd}$ is a smoothly increasing function of $m_P^{-1}$; thus the static
result may be a lower bound for $r_{sd}$.

The extraction of the individual values of $\MB$ and $\MS$ is clearly expected
to have larger errors. Thus, conventionally~\cite{CB88,FLYNN,SONI} these matrix
elements are given in terms of the corresponding B parameter,
which is better behaved. 
Carrying out a similar continuum extrapolation as above
for $B_{bd}(\mu)$, we find a constant fit yields 
$B_{bd}(2 \,{\rm GeV})=0.97(3)$ while 
linear extrapolation in $a$ gives 1.02(13). We cannot, however,
distinguish $B_{bs}(2 \,{\rm GeV})$ 
from $B_{bd}(2 \,{\rm GeV})$ since our data for $B_{hl}$ {\em vs.} $\kl^{-1}$ 
are fit equally well to constant or linear fits.
This was not true for $\MP$, as is evident from Fig.~\ref{mllc}.
Using linear extrapolations in both $a$ and $\kl^{-1}$, 
we quote $B_{bd}(2 \,{\rm GeV})=B_{bs}(2 \,{\rm GeV})=1.02(13)$,
where the error is purely statistical (systematic errors are small
in comparison). 

\begin{figure}[hbt]
    \vbox{
        \epsfxsize=6.0in \epsfbox[0 0 4096 4096]{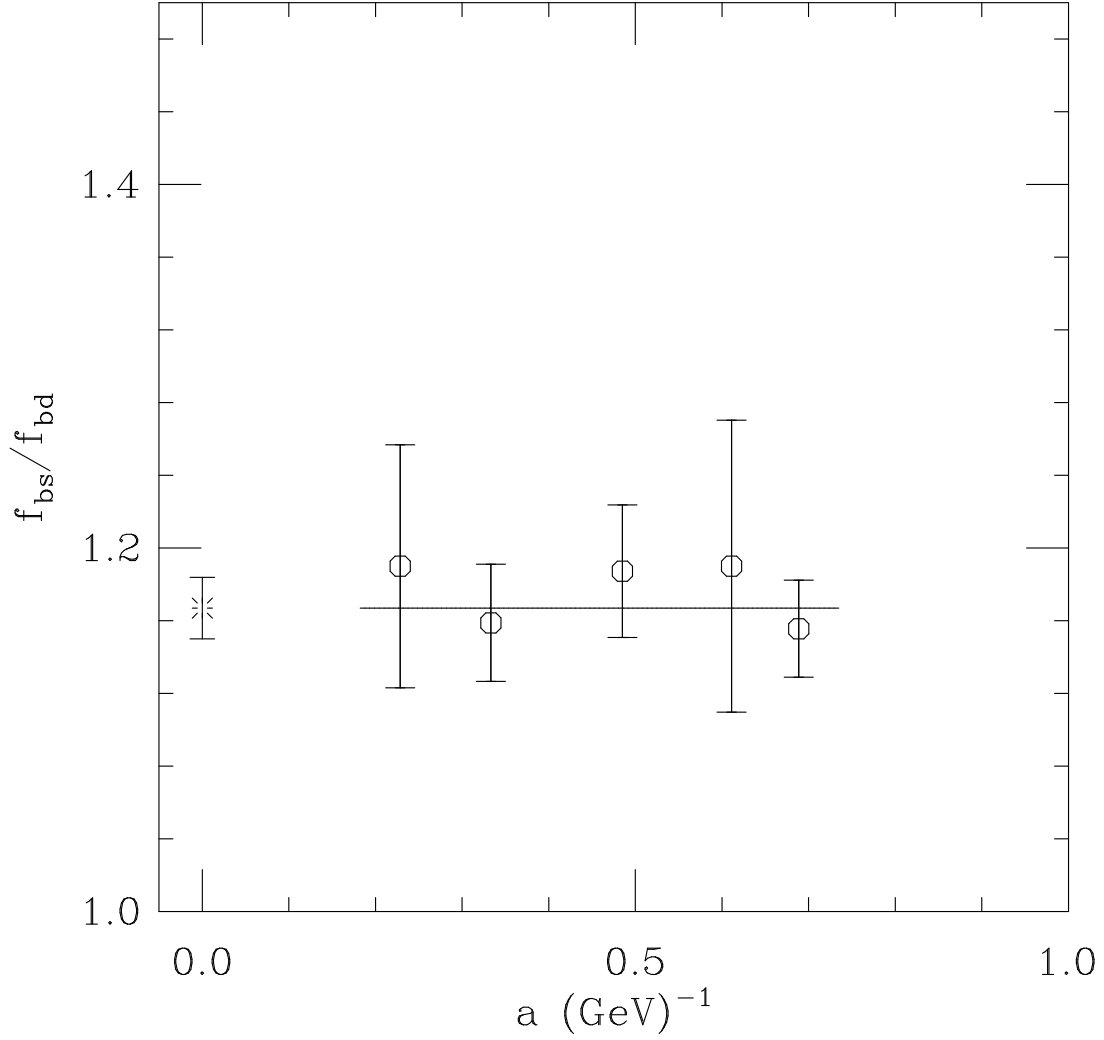}
    }
       \caption{\label{decay ratio} The SU(3) flavor breaking ratio of decay constants
	$f_{bs}/f_{bd}$ {\em vs.} the lattice spacing $a$. Results are plotted for
	linear chiral extrapolations.
	The line denotes a constant fit, and the burst is the corresponding
	continuum extrapolation.}
\end{figure}

We recall that until now~\cite{FLYNN,SONI}, lattice results for the SU(3) 
breaking ratio $r_{sd}$ have been obtained by using Eqs.~(\ref{ratio}) 
and (\ref{B param})
and the lattice determinations of $f_{bd(s)}$ and $B_{bd(s)}$.
A simultaneous fit of
the pseudoscalar and axial vector
correlators yields the decay constant $f_{hl}$.
Using Eq.~\ref{heavy decay} plus corrections up to $O(m_{hl}^{-2})$,
we find for the ratio of $B$ meson
decay constants, $f_{bs}/f_{bd}=1.17(2)(+2)(+5)(+1)(-2)(\pm 4)$
(see Fig.~\ref{decay ratio} and Table~\ref{r_decay table}). This result is for a constant
continuum extrapolation which is reasonable for the data
shown in Fig.~\ref{decay ratio}. The uncertainties are
statistical,  and the following systematic differences from:
(2.) using the pole mass instead of the kinetic mass,
(3.) using quadratic chiral extrapolations for the heavy-light mesons 
instead of the linear ones,
(4.) replacing quadratic chiral extrapolations with linear
extrapolations in the determination of  $\kappa_c$ and
$\kappa_s$, and 
(5.) a shift in $k_c$ as before.
Again, the numbers in parentheses correspond to the columns in Table~\ref{r_decay table}.
Finally, we consider
the overall
uncertainty in the slope of the ratio versus lattice spacing.
A linear continuum extrapolation using all the
data has a negative slope; while omitting the point at $\gbeta=5.7$
yields a positive slope. Also, the results in columns 4 and 5 have 
positive slopes. 
Thus we include a symmetric error of $\pm0.04$ due to the continuum 
extrapolation.
For $f_{bs}/f_{bd}$,
there were no significant differences due to changing the fit range by one
or two units. Adding the above systematic errors,
we find $f_{bs}/f_{bd}=\RDECAY$.
We have omitted the larger volume at $\gbeta=6.0$ from the analysis for 
reasons similar to those
described above.  In addition, the data set at $\gbeta=5.85$ causes 
the same difficulties 
as before. 
For the ratio of decay constants we are able
to get a statistically significant result at $\gbeta=6.5$ which is included in
the above analysis.

\begin{table*}
\caption{Summary of results for the ratio $f_{bs}/f_{bd}$.
The last two rows refer to constant and linear continuum
extrapolations, respectively. 
The errors shown in parentheses are statistical.
Columns 2-5 represent systematic differences relative to column 1 
(see text). Where there is no entry, a reasonable fit was not found.}
\label{r_decay table}
\begin{tabular}{cccccc}
$\gbeta$&1&2&3&4&5 \\
\hline
5.7&		1.156 (26)&1.165 (97) &1.151 (66) &1.181 (30)&1.140 (25)\\
5.85&	   	1.190 (80)&-          &1.493 (249)&1.184 (79)&1.287 (95)\\
6.0$(16^3)$&	1.187 (36)&1.180 (92) &1.204 (61) &1.182 (34)&1.172 (35)\\
6.3&		1.159 (32)&1.180 (93) &1.266 (74) &1.177 (33)&1.149 (31)\\
6.5&		1.190 (66)&1.289 (192)&1.408 (243)&1.145 (40)&1.100 (46)\\
$\infty$&	1.167 (17)&1.184 (52) &1.214 (37) &1.174 (17)&1.148 (16)\\
$\infty$&	1.183 (55)&1.241 (169)&1.375 (138)&1.152 (45)&1.133 (44)\\
\end{tabular}
\end{table*}

Our result for $f_{bs}/f_{bd}$ 
is consistent with previous estimates~\cite{FLYNN,SONI,BLS,UKQCD}.
Note that while the decay constant using Wilson quarks has a perturbative
correction (which does not depend on the scale $\mu$), 
it cancels in the ratio (up to small quark mass corrections).
As indicated above, the ratio of B parameters is consistent with unity, and
the ratio of masses is 1.017\cite{PDB},
so the old method leads to $r_{sd}\approx 1.42(5)^{+28}_{-15}$
which is compatible with, though somewhat lower than,
$\CV$ from our direct method.
As we have emphasized, the direct method has many
desirable features which may allow future lattice computations 
to significantly improve the
precision of this method for the determination of the ratio $r_{sd}$.

This research was supported by US DOE grants DE-FG02-91ER4D628 
and DE-AC02-76CH0016.
The numerical computations were carried out at NERSC. 
We thank Jonathan Flynn and Guido Martinelli for pointing 
out an error in the equation
corresponding to (5) in an earlier version of this work \cite{US96}.

\newpage

\newpage
\newpage
\newpage
\newpage


\begin{thebibliography}{-9}
\bibitem{BEES}We will use the interchangeable notation $B^0$ or 
$B_{bd}$ for B mesons that are bound states of $\bar b d$; similarly
$B^0_s \equiv\ B_{bs}$.
\end{thebibliography}
\end{document}